# Two-Stage Coded Distributed Edge Learning: A Dynamic Partial Gradient Coding Perspective

Tingting Yang, *Member*, *IEEE*, Xinghan Wang, Jiahong Ning, Yang Yang, *Fellow IEEE*

*Abstract*—Federated edge learning (FEL) can training a global model from terminal nodes' local dataset, which can make full use of the computing resources of terminal nodes and performs more extensive and efficient machine learning on terminal nodes with protecting user information requirements. Performance of FEL will be suffered from long delay or fault decision as the master collects partial gradients from stragglers which cannot return correct results within a deadline. Inspired by this, in this paper, we propose a novel coded FEL to mitigate stragglers for synchronous gradient with a two-stage dynamic scheme, where we start with part of workers for a duration of before starting the second stage, and on completion of at the first stage, we start remaining workers in the second stage. In particular, the computation latency and transmission latency is essential and should be quantitatively analyzed. Then the dynamically coded coefficients scheme is proposed which is based on historical information including worker completion time. For performance optimization of FEL, a Lyapunov function is designed to maximize admission data balancing fairness and two stage dynamic coding scheme is designed to maximize arrival data among workers. Experimental evidence verifies the derived properties and demonstrates that our proposed solution achieves a better performance for practical network parameters and benchmark datasets in terms of accuracy and resource utilization in the FEL system.

## I. INTRODUCTION

With the improvement of mobile device performance and the development of edge networks, the concept of federated learning can be flexibly applied to edge computing scenarios. Because a large number of computing resources are configured in the network at the edge, a centralized data center is no longer necessary for federated learning. Model learning and data training can be done in a distributed manner, and the number of uploads is avoided. Network load and latency caused by huge raw data. At the same time, because federated learning has the advantage of utilizing real data from a large area of user clusters, it greatly facilitates unprecedented large-scale and flexible data training and model learning. Based on the above advantages, wireless federated edge learning has attracted the attention of many people in the industry.

This work was supported by the National Key R&D Program of China under Grant 2020YFB1806800, Guangdong Province Basic and Applied Basic Research Foundation under Grant 2019B1515120084, and Key Projects of the National Defense Foundation Strengthening Plan under Grant 2020-JCJQ-ZD-020-05, the Peng Cheng Laboratory Project (Grant No. PCL2021A02)

According to the training method of federated learning, training algorithms can be divided into *synchronous training* [1-2] *algorithms*, *asynchronous training algorithms* [3-4] and *hybrid training algorithms* [5-8]. In synchronous training, scheduling clients and updating models are synchronously blocked. Each server update needs to receive feedback from enough clients, as shown only waiting for all clients (nodes) to train the model only execute the update global model, and then download it to the client. The asynchronous training method divides the scheduling client and the update model into parallel threads. The training model of each client (node) can be uploaded in parallel to update the global model without waiting for all clients. After the client completes the global update, it can schedule the client periodically, and the server can update the global model after receiving the model parameters uploaded by any client. Hybrid federated learning combines the two methods of synchronization and asynchronous, it can enforce synchronization in a local scope, wait until the training model of some clients. The asynchronous updates are uploaded to the server side together, and then downloaded to the client side after the global update is completed.

Ma et al. [5] proposed a semi-asynchronous federated learning mechanism (FedSA), in which the parameter server selects a certain number of local models for aggregation based on the order in which local models arrive in each round. In [6] they proposed a joint asynchronous and synchronous federated edge learning architecture, SAFA. Before a round of local training, the server classifies the client into three states: up-to-date, deprecated, and tolerable, based on the model version currently held by the client. The latest and deprecated clients are forced to update to the latest global model (forced sync), while tolerable clients can continue to use their previous local model results to some extent (tolerate some asynchrony). Zhou et al. [7] grouped the related workers into a community and perform synchronous updates in the same community, which can speed up the training aggregation process, called community-aware parallel synchronization. The asynchronous actor-critic algorithm intelligently determines community deployment. Wang et al. [8] proposed an efficient federated edge learning mechanism, which divides edge nodes into *K* groups through balanced clustering. The edge nodes in the cluster forward their local updates to the cluster header to aggregate in a synchronous manner, called cluster aggregation, while all clusters perform an asynchronous global aggregation method.

However, in the synchronization part of the hybrid federated edge learning, the parameter server has to wait for the slowest client. Due to the heterogeneity of the edge, the waiting time of idle participants is too long. The asynchronous part requires frequent model transmission, resulting in a lot of communication resource consumption. In addition, the frequency of different clients participating in asynchronous updates may seriously affect the accuracy of training, especially in terms of transmission efficiency and resource utilization, which are not well guaranteed.

Inspired by these, in this paper, we propose a novel coded FEL to mitigate stragglers for synchronous gradient with a two-stage dynamic scheme to enhance accuracy and resource utilization in the FEL system. The contributions of this article are listed as follows:
1) We propose a novel coded FEL to mitigate stragglers for synchronous gradient with a two-stage dynamic scheme, where we start with part of workers for a duration of before starting the second stage, and on completion of at the first stage, we start remaining workers in the second stage.
2) In the proposed scheme, the computation latency and transmission latency is essential and quantitatively analyzed. Then the dynamically coded coefficients scheme is proposed which is based on historical information including worker completion time. For performance optimization of FEL, a Lyapunov function is designed to maximize admission data balancing fairness and two stage dynamic coding scheme is designed to maximize arrival data among workers.
3) The extensive experiments demonstrate the effectiveness of the proposed scheme, in which can achieve the performance improvement by reducing the computation time and transmission time while keeping the convergence rate.

## II. RELATED WORKD

With the improvement of mobile device performance and the development of edge networks, the concept of federated learning can be flexibly applied to edge computing scenarios. Because a large number of computing resources are configured in the network at the edge, a centralized data center is no longer necessary for federated learning. Model learning and data training can be done in a distributed manner, and the number of uploads is avoided. Network load and latency caused by huge raw data. At the same time, because federated learning has the advantage of utilizing real data from a large area of user clusters, it greatly facilitates unprecedented large-scale and flexible data training and model learning. Based on the above advantages, wireless federated edge learning has attracted the attention of many people in the industry. At present, research is mainly carried out from the following aspects: the efficiency and effectiveness of wireless federated edge learning [1], the security of data protection [2-3], Robustness in the face of malicious attacks [4], overall reliability and fairness of the system [5] and so on. Among them, reducing communication overhead to improve the efficiency and utility of wireless federated edge learning is very important in the current edge network environment, which has attracted the interest of many researchers.

According to the training method of federated learning, training algorithms can be divided into *synchronous training algorithms*, *asynchronous training algorithms* and *hybrid training algorithms*. In synchronous training, scheduling clients and updating models are synchronously blocked. Each server update needs to receive feedback from enough clients, as shown in Figure 1-a, only waiting for K clients (nodes) to train the model Only execute the update global model, and then download it to the client. The asynchronous training method divides the scheduling client and the update model into parallel threads. As shown in Figure 1-b, the training model of each client (node) can be uploaded in parallel to update the global model without waiting for K clients. After the client completes the global update, it can schedule the client periodically, and the server can update the global model after receiving the model parameters uploaded by any client. Hybrid federated learning combines the two methods of synchronization and asynchronous, as shown in Figure 1-c, it can enforce synchronization in a local scope, as shown in Figure 1-c, wait until the training model of 3 clients and other local The asynchronous updates are uploaded to the server side together, and then downloaded to the client side after the global update is completed.

Although many successful experiences of federated learning have used synchronization algorithm training, because the server and client are synchronized, it will cause communication congestion and lead to the phenomenon of stragglers. Therefore, the synchronization algorithm is still not flexible enough on the scheduling client, and the synchronization overhead is high. Some studies have begun to combine asynchronous algorithms with federated learning. The server regularly sends the latest model to the client, but does not block and waits for the client to reply, and then updates the global model immediately after receiving the update from the client. However, the asynchronous algorithm has the problem of staleness caused by the delay of model uploading. Some clients are still training the old version of the global model, but the server has iterated the new global model using other clients' models. The version difference generated in this case has a negative impact on the training accuracy, especially when the server receives a large number of high-concurrency updates at the same time, due to the acceleration of the global model iteration, the version difference will intensify, which will cause more serious negative effects. Based on the advantages and disadvantages of the above synchronization and asynchronous, researchers combine the two to reduce the consumption of network resources and improve the learning efficiency.

### A. Synchronization federated learning

Early wireless federated edge learning mainly used synchronous training. In [6], McMahan proposed the FedAvg algorithm, which employs synchronous communication to schedule clients, and in each round, the server will select some clients to download the current global model. The client will then use this global model to train a new local model on the local data and send the trained model parameters to the server.

Finally, the server should receive the model parameters of all clients selected for this round, then compute a new global model, and then select a new set of clients for the next round of training. However, in each round, some clients will fail to train due to poor network and other sudden problems. Therefore, if the server does not receive enough updates, the server will abandon this round of updating the global model.

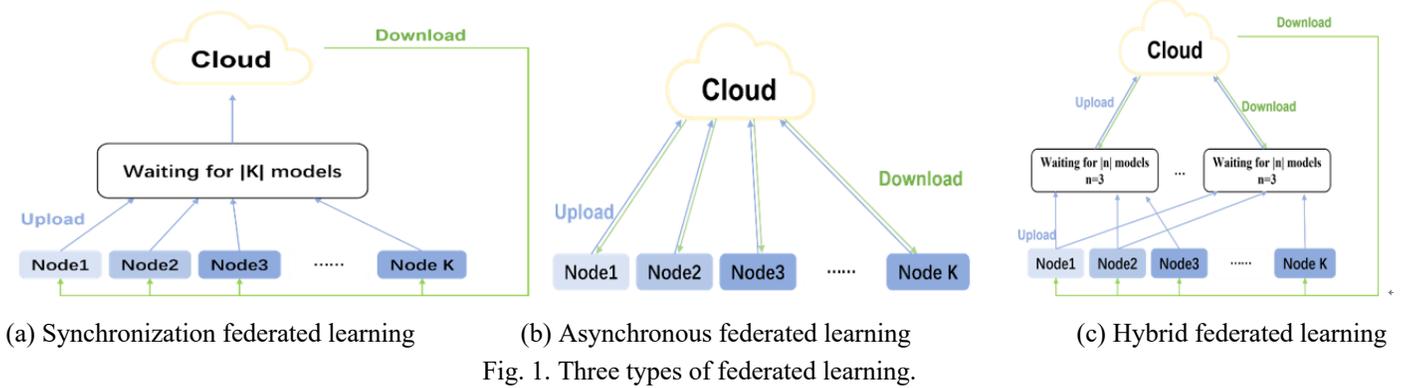

(a) Synchronization federated learning  (b) Asynchronous federated learning  (c) Hybrid federated learning

Fig. 1. Three types of federated learning.

In federated edge learning, due to the diversity of devices and the nature of loose federation, synchronization algorithms typically face two challenges when scheduling distributed training: (1) Unpredictable client situations. A client-based distributed network is not a highly controlled environment, so the server cannot know exactly whether the client's conditions are suitable for participating in training. Even if the client has notified the server that it is ready to run the training task at the start of training, the client may suddenly experience a power failure, disconnection, or other emergency during training, causing it to lose contact with the server. (2) The communication frequency is unstable. Because clients are in a complex and uncontrolled environment, they cannot always maintain a running state of an executable task. Therefore, the server determines whether to schedule tasks for the client through the communication protocol. This creates a phenomenon where communication is frequent in good situations and infrequent in bad situations. But synchronous training is more suitable for frequent communication.

**B. Asynchronous federated learning**

Compared with synchronous update, asynchronous update has the following advantages: when the gradient update delay (version difference) is small, the convergence speed of asynchronous communication is much faster than that of synchronous update; since the server does not wait for the client's response in asynchronous training, the server can More flexibility in coordinating client participation in training; asynchronous updates can run more clients in parallel compared to synchronous updates, which can avoid network congestion. However, the asynchronous update algorithm has the problem of staleness caused by model upload delay. Some clients are still training the old version of the global model, but the server has iterated the new global model using other clients' models. The version difference generated in this case has a negative impact on the training accuracy, especially when the server receives a large number of high-concurrency updates at the same time, due to the acceleration of the global model iteration, the version difference will intensify, which will cause more serious negative effects.

In synchronous federated edge learning, the parameter server waits for the slowest client, resulting in excessively long waiting times for idle participants due to edge heterogeneity. On the other hand, although asynchronous federated edge learning can well address edge heterogeneity, it requires frequent model transfers, resulting in a large consumption of communication resources. Furthermore, the frequency with which different clients participate in asynchronous updates can seriously affect training accuracy, especially on independently identically distributed (Non-IID) data. To address these issues, researchers recently proposed a hybrid/semi-asynchronous federated edge learning architecture.

In response to the above problems, researchers propose a series of federated edge learning strategies based on asynchronous training to improve speed and scalability. Xie et al. [26] proposed a new asynchronous federated optimization algorithm, the key idea of which is to use a weighted average to update the global model, which can adaptively set the mixed weight value according to the staleness function. To allow each edge device to use its local data more efficiently, Lee et al. [27] proposed an asynchronous federated wireless distributed learning algorithm architecture, emphasizing that asynchronous transmission scheduling should be carefully determined to account for system uncertainties. In [28], they proposed an Adaptive Asynchronous Federated Learning (AAFL) mechanism to address the problem of high resource cost due to edge data imbalance, edge device dynamics, and resource constraints. To handle the dynamic nature of edge devices, local updates are aggregated at a ratio α according to the order in which they arrive at the parameter server in each epoch. In addition, the mechanism can dynamically select local updates for global model aggregation under different epoch network conditions. Considering the problem of long training time and high bandwidth cost, based on the idea of local updates aggregated at a certain ratio $\alpha$, they proposed in [28], in [29], they proposed an asynchronous federated edge learning strategy for bandwidth allocation optimization. Wang et al. [30] proposed a new asynchronous federated edge learning architecture that takes into account potential failures when local model updates are uploaded and staleness between globally updated models of varying degrees produced during the update process.

## C. Hybrid federated learning

Ma et al. [32] proposed a semi-asynchronous federated learning mechanism (FedSA), in which the parameter server selects a certain number of local models for aggregation based on the order in which local models arrive in each round. They theoretically analyze the quantitative relationship between the convergence bound of FedSA and different factors such as the number of participating users per round, the degree of data non-IID and marginal heterogeneity. In [33] they proposed a joint asynchronous and synchronous federated edge learning architecture, SAFA. Before a round of local training, the server classifies the client into three states: up-to-date, deprecated, and tolerable, based on the model version currently held by the client. The latest and deprecated clients are forced to update to the latest global model (forced sync), while tolerable clients can continue to use their previous local model results to some extent (tolerate some asynchrony). This mitigates the effects of straggling, crashes and model obsolescence, improving efficiency and quality. In [34], they proposed a hierarchical federated edge learning architecture for HPC, FedAT. The clients are layered according to the submission speed of each client. The more training is performed, the smaller the weight of a single update is, that is, the influence of the layer models with slow updates on the global model is guaranteed. Only model training at the same layer is synchronous, and model training at different layers is asynchronous. In the Internet of Vehicles (IoV) system, due to the high mobility and uncertainty of vehicles, the existing federated learning protocols are difficult to meet all the requirements of the IoV system, such as efficient resource allocation, high-precision learning, and fast learning algorithms. To address the above problems, Liang et al. [35] proposed a Semi-Synchronous Federated Learning (Semi-SynFed) protocol to improve the performance of machine learning in the Internet of Vehicles. In Semi-SynFed, the training samples are firstly selected according to their computing power, network capability and learning value, and appropriate nodes are selected to participate in the aggregation.

The research on heterogeneity in current research work mainly prioritizes fast workers and reduces the participation of slow workers, which in turn leads to workload imbalance and computational inefficiency. In response to this problem, Zhou et al. [36] grouped the related workers into a community and perform synchronous updates in the same community, which can speed up the training aggregation process, called community-aware parallel synchronization. The asynchronous actor-critic algorithm intelligently determines community deployment. Aiming at the problem that the existing federated learning mechanism may lead to long training time and consume a lot of communication resources, Wang et al. [37] proposed an efficient federated edge learning mechanism, which divides edge nodes into $K$ groups through balanced clustering. The edge nodes in the cluster forward their local updates to the cluster header to aggregate in a synchronous manner, called cluster aggregation, while all clusters perform an asynchronous global aggregation method.

However, in the synchronization part of the hybrid federated edge learning, the parameter server has to wait for the slowest client. Due to the heterogeneity of the edge, the waiting time of idle participants is too long. The asynchronous part requires frequent model transmission, resulting in a lot of communication resource consumption. In addition, the frequency of different clients participating in asynchronous updates may seriously affect the accuracy of training, especially in terms of transmission efficiency and resource utilization, which are not well guaranteed.

## III. SYSTEM MODEL AND PROBLEM FORMULATION

In this section, we will describe the system model and use of the coding theory to remove the bottlenecks caused by the data placement phases, data delivery phases, gradient computation phases and model uploading phases, for the proposed the tow-stage dynamic coded federated learning (TSDCFL) in MEC.

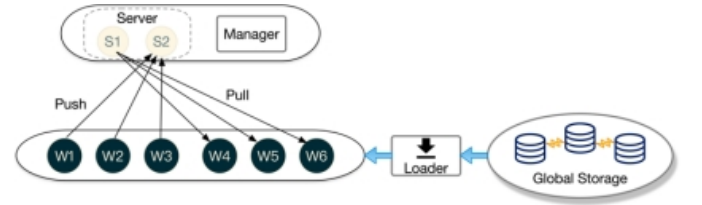

Fig. 1. TSDCDL architecture

We consider a FEL system with the parameter edge servers, i.e., model owners, each of which aims to develop a model. The set of local workers denoted as $M = \{1,2,3...,M\}$, i.e., worker that have the computation and communication capabilities can support the model owners. The data owner that can associate themselves with workers in order to deliver data to workers. As shown in Fig.1.

### 1. Gradient coding

Given the data set $D$ is divided into $K$ non-overlapping equal-size types, denoted by $D = \{D_1,...,D_K\}$. Let $g_k(i)$ denotes the partial gradient over a data partition $D_k$ in the epoch $i$ and can be calculated as:

$$g_{k,<i>} = \frac{1}{|D_k|} \sum_{(x,y) \in D_k} \nabla l(x,y) \quad (1)$$

Then, we note that the whole aggregated gradient is given by $g_{<i>} = \sum_{k=1}^{K} g_{k,<i>}$. The parameters server is update by:

$$w_{<i+1>} = w_{<i>} - \eta g_{<i>} \qquad (2)$$

If the data partition is assigned to the worker, each worker computes the corresponding partial gradient, and the parameter server aggregate the partial gradient. Due to the limited by workers local computation capacity, limited harvested energy and high transmission delay, there exists some straggler. At the each iteration, the data loader assigns redundancy partial gradients to workers, thus, each worker is responsible for coded partial gradient so that parameter server can recover the whole gradient from subset of workers when the machine break down or the partial gradient cannot be transmitted to server. As shown in Fig.2. In each round, the gradient coding can tolerate up to $s_{<i>}$ stragglers. Formally, there exists a set of the non-straggler $M_{non\_stragglers} \in M$ with $M_{non\_stragglers} = M - s_{<i>}$. Thus, in the epoch $i$, the whole aggregated gradient can be recovered from any $M - s(i)$ code words.

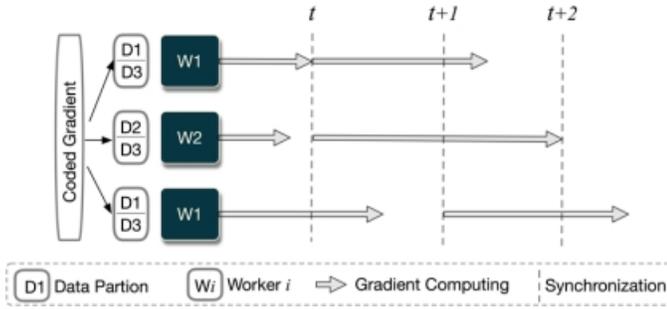

Fig. 2. The cooperative federated learning weight update

## 2. Tow stage gradient coding

Since the complexity of encoding and decoding is related to the number of workers and the number of data partitions, if we use encoding and put a lot of redundancy on each worker at the beginning, the data partitions will be repeated by multiple workers, resulting in a lot of consumption and time delay. We use two-stage gradient coding, where we start with $M_1$ out of $M$ workers for a duration of $T_{comp,<i>}$ before starting the second stage, and on completion of $M_c$ at the first stage, we start $M - M_1$ workers. As shown in Fig. 3.

In the tow stage gradient coding, we consider two stages. Initially, the data loader assigns the data partitions to first $M_1$ workers and $M_1$ workers compute the partial gradients from the data partitions,. At the first stage, $M_1$ workers start computing the partial gradients and $M_c$ out of $M_1$ workers complete the gradient computation for a duration of $T_{comp,<i>}$. At the second stage, $M - M_1$ workers start computing the partial gradients assigned to them based on which $M_c$ out of $M_1$ workers have finished. Tow stage gradient coding provide the flexibility that the gradient computation is complete when any $M_{non\_stragglers}$ out of $M$ finish their assigned data partitions. The aim is that as long as $M_c$ can execute their assigned data partitions at the first stage and $M_{non\_stragglers} - M_c$ can execute their assigned data partitions at the second stage, the whole gradients computation gets completed. Tow stage gradient coding exploits the flexibility that not all the workers are started the same time, which helps to reduce the gradients computation time and total workers utilization. The type of data partitions is lower than starting all $K$ types compared other coding scheme, which can reduce the complexity of encoding and decoding.

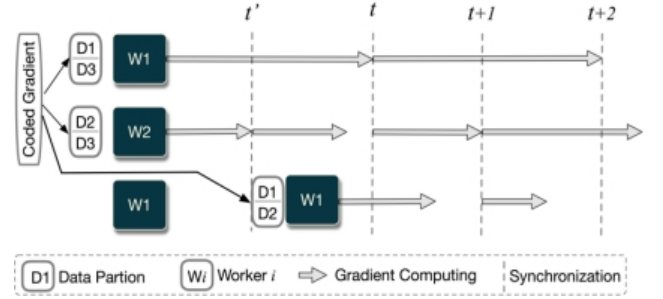

Fig. 3. The cooperative federated learning weight update

We seek tow stage gradient coding scheme that are robust to any $s_{<i>}$ stragglers in each epoch $i$.

**Lemma1.** *Tow stage gradient coding strategy is robust to any $s(i)$ stragglers in each epoch $i$ if $B_1$ and $B_2$ satisfy the following span condition, in where concludes tow metrics $B_1 = [b_1, b_2, ..., b_{M_1}]^T \in \mathbb{R}^{M_1 \times K}$ at the first stage and $B_2 = [b_1, b_2, ..., b_{M-M_1}]^T \in \mathbb{R}^{(M-M_1) \times K}$ at the second stage.*

(**Span Condition**): *for any subset $I_1 \in [M_c]$, $I_2 \in [M_1 - M_c]$ and $I_3 \in [M - M_1]$, such that $|I_1 \cup I_2 \cup I_3| = M - s_{<i>}$,*

$$1_{1 \times K} \in span\{b_i, b_j \mid i \in I_1 \cup I_2, j \in I_3\},$$

where $1_{1 \times K}$ is an all one vector, $span()$ is the span of vectors, and $[]$ denotes the set.

Let decoding strategy $A \in \mathbb{R}^{S \times M}$ denote the all $S$ stragglers patterns, where $S = \binom{M - M_c}{s_{<i>}}$ represents the number of possible straggler which can be tolerated by tow stage gradient coding. Columns of $A$ are indexed by the workers. The rows of the $A$ are denotes by $a_j$. Each $a_j$ denotes a specific scenario of stragglers and zeros in $s(i)$ out of the $M - M_c$ positions.

Hence, the decoding strategy can be constructed by:

$$A \begin{bmatrix} B_1 \\ B_2 \end{bmatrix} = 1_{S \times K} \qquad (3)$$

We consider one row of equation and decoding function can also be a linear combination as:

$$a_j \left| \sup p(a_j) \begin{bmatrix} B_1 | I_1 \cup I_2 \\ B_2 | I_3 \end{bmatrix} \right. = 1_{1 \times K} \qquad (4)$$

## 3. Latency Analysis

As mentioned before, we aim at reducing complexity of encoding and decoding and the local gradient computation latency. Stragglers are inevitable in distributed learning systems, due to various reasons, including network latency, transmission fair among all workers, channel state, and workload imbalance. In order to mitigate the adversarial effects of stragglers, we mitigate the impact from perspective of reducing computation time and transmission time. We dynamically adjust the coded coefficients based on historical information including worker completion time. Therefore, the computation latency and transmission latency is essential and should be quantitatively analyzed. We aim to reduce the number of stragglers in each epoch and ensure the fairness of the transmission among all workers.

We consider that the entire training process is composed of multiple time slots, and maximum length of time slot is $T$. The system operate in a time slotted structure with length $T$, which means channels remain unchanged during a time slot and vary between slots. Let $v_{m,<i>}(t)$ denote the worker $m$ transmission time and not exceed the slot length, given as

$$0 \leq v_{m,<i>}(t) \leq T \qquad (5)$$

And the total available transmission time of workers cannot exceed the available sub-channels time and a worker can be assigned more than one channels at time slot t, given as

$$\sum_{m \in M_{non\_strgglers,<i>}} v_{m,<i>}(t) \leq TL(t)$$
$$v_{m,<i>}(t) = \sum_{j=1}^{L(t)} v_{m,<i>}^j(t) \qquad (6)$$

where $L(t)$ denotes the available channels in time slot $t$ and the transmission amount of data is determined by the backlog of data and channel capacity, given as

$$c_{m,<i>}(t) = \min\{Q_{m,<i>}(t), r_{m,<i>}(t)v_{m,<i>}(t)\}$$

where $r_m(t)$ is the channel capacity of worker $m$ during time slot and $Q_m(t)$ denote the backlog of data from worker $m$, $Q_m(t)$ is updated along the time, as given by

$$Q_{m,<i>}(t+1) = Q_{m,<i>}(t) + d_{m,<i>}(t) - c_{m,<i>}(t) \qquad (7)$$

where $d_m(t)$ is admission data from worker $m$, this part of data come from the gradient vector obtained by the worker performing backward propagation and accumulate on worker over time. Worker may only be able to admit part of the arrived data, as given by

$$0 \leq d_{m,<i>}(t) \leq D_{m,<i>}(t) \qquad (8)$$

where $D_m(t)$ is the arrival data from worker $m$. The energy consumption of the worker consist of two parts: One part comes from the uploading parameter server, and the other part comes from gradient computation. Let $e_{m\_up,<i>}(t)$ denotes the energy consumption from uploading data and $e_{m\_com,<i>}(t)$ denotes the energy consumption from gradient computation, can be written as

$$e_{m\_up,<i>}(t) = p_m v_{m,<i>}(t) \qquad (9)$$

$$e_{m\_com,<i>}(t) = f_m(t)\delta_m \qquad (10)$$

where $p_m$ is the transmit power of worker $m$, $f_m$ specifies the required CPU cycles to calculate the gradient from worker $m$ and $\delta_m$ is the coefficient energy consumption required per CPU cycle. The energy harvesting from worker can be consider as a stochastic process. Assume each worker can harvest at most $E_{m,<i>}^H(t)$ with the maximum $E_{\max,<i>}^H(t)$ among whole workers, and each worker can store part of the newly harvested energy, denotes by $e_{m\_store,<i>}$, where

$$0 \leq e_{m\_store,<i>} \leq E_{m,<i>}^H(t) \leq E_{\max,<i>}^H(t).$$

Let $E_{m,<i>}(t)$ denote the battery backlog of worker $m$, can be written as

$$E_{m,<i>}(t+1) = E_{m,<i>}(t) - e_{m\_up,<i>}(t) - e_{m\_com,<i>}(t) + e_{m\_store,<i>} \qquad (11)$$

Let $R_m(t)$ denotes the required CPU cycles to process gradient computation, can be updated by

$$R_{m,<i>}(t+1) = \max[R_{m,<i>}(t) - f_m(t), 0] \qquad (12)$$

Upon the receipt of gradient from non-stragglers worker in parameter server, the parameter server will update the parameter. Let $R_{server}(t)$ be the required CPU cycles to update parameters at the server, and can be updated by

$$R_{server,<i>}(t+1) = \max[R_{server,<i>}(t) - F(t), 0] + \sum_{m \in M_{non\_stragglers}} c_{m,<i>}(t)\xi_m \qquad (13)$$

where $\xi_m$ is the number of CPU cycles required per bit of the worker $m$.

## 4. Problem Formulation

Due to the existence of stragglers, we must consider the encoding method to reduce the impact of stragglers to ensure that all data partitions can be updated in each epoch. But if the encoding start at the beginning, it will increase complexity of encoding and decoding and increase the burden on each worker, so we first start to calculate the gradient from a part of the workers within the deadline, then we need to count the remaining data partitions, and then predict the stragglers based on the historical status and the historical completion time of each worker, which can improve the utilization of the worker and speed up the time of each epoch. The purpose of adding coding is to ensure that all data partitions can be updated in each epoch, then we need to reduce the time of each epoch by reducing the number of stragglers as much as possible, increasing the number of local data generated by workers, balancing the network throughput and fairness among the workers. We reduce the computation time by maximizing the arrival data, and reduce the transmission delay and ensure the fairness of the transmission by maximizing the admission data. Specifically, the increase in arrival data can reduce the training time per epoch, which means that the higher the arrival data, the less likely the worker becomes a straggler is in the computation phase. Admission data is related to throughput and fairness, and increasing admission data can reduce the transmission time. Besides the tolerance of stragglers in computation phase, we first consider about arrival data from each worker. The arrival data of epoch $i$ for a given data assignment depends on the straggler prediction and the two stage gradient coding strategy. Here, our objective is to maximize the expected arrival data based on the past straggler behavior:

$$\text{P1: max } \mathrm{E}_{s_{<i>}|s(i-1)}\left[D(\tau_{<i>}, s_{<i>}, \mathrm{B}_1, \mathrm{B}_2)\right] \quad (14)$$
$$\text{s.t. } \mathrm{B}_1 \text{ and } \mathrm{B}_2 \text{ satisfy Span Condition}$$

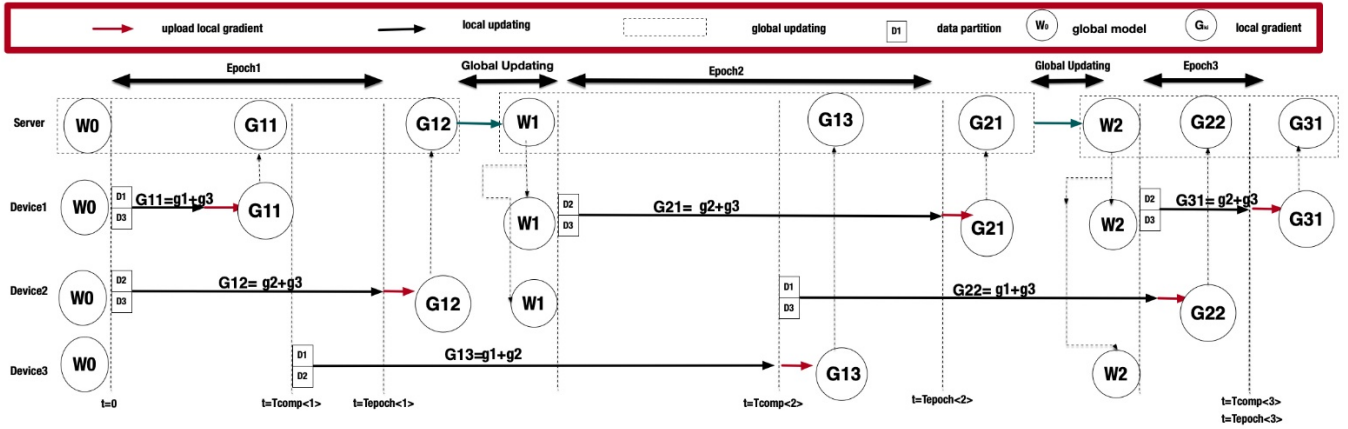

Fig. 4. Two-stage coding scheme, TSDCFL.

By solving the **P1** is able to maximize arrival data, and enable to eliminate stragglers in computation phase.

Once the worker finishes training in the last iterations of each epoch, the worker starts uploading gradient vector to the parameter server, and the upload size is related to the admission data. We need to maximize the throughput and guarantee fairness in distributed learning system, so as to reduce the likely the worker becomes a straggler in the transmission phase. **P2** begins to optimize the admission data by getting arrival data from **P1**. Leveraging between throughput and fairness in system, the objective **P2** can be defined as:

$$\text{P2: max } \log(1+\lambda_m \bar{d}_m)$$
$$\text{s.t. C1: } 0 \leq \upsilon_{m,<i>}(t) \leq T$$
$$\text{C2: } 0 \leq d_{m,<i>}(t) \leq D_{m,<i>}(t)$$
$$\text{C3: } 0 \leq e_{m\_store,<i>} \leq E^H_{m,<i>}(t) \quad (15)$$
$$\text{C4: } e_{m\_up,<i>}(t) + e_{m\_com,<i>}(t) \leq E_{m,<i>}(t)$$
$$\text{C5: } \bar{Q}_{m,<i>}, \bar{E}_{m,<i>}, \bar{R}_{m,<i>} \text{ and } \bar{R}_{server,<i>} < \infty$$

where $\bar{X} = \lim_{T \to \infty} \frac{1}{T}\sum_{t=0}^{T} \mathrm{E}[X(t)]$ and the total admission data is equal to system throughput, $\lambda_m$ is a positive constant to characterize the weight of the admission data.

### 4.1 Convergence Analysis

We make the following assumptions:
**Assumption 1.**
1) *Lipschitzian Gradient: The gradient function is Lipschitz continuous with constant L>0. Equivalently, for two parameter , the gradient function satisfies:* $\|g(w) - g(\bar{w})\| \leq L\|w - \bar{w}\|$.

2) *Unbiase gradient: for all parameter, the computed partial gradient is unbiased* $\mathrm{E}[g_k(w_{<i>})] = g_k(w_{<i>})$.

By denoting the minimum loss function, we establish the following theorem for the convergence of TSDCFL.
**Theorem 1**. *Bounded variance: there exist scalar such that*

$$\frac{1}{P}\sum_{l=1}^{P}\|g(w_{<l>})\|^2$$
$$\leq \frac{2}{\bar{\eta}P}(F(w_{<1>}) - F(w_*)) + \frac{3L\bar{\eta}}{K}(1+\bar{\eta}\tau_{\max})m(C_1+C_2)\varsigma^2$$

where C1 and C2 are the maximum of coding coefficient and decoding coefficient in first stage and second stage, $n$ is the step size, $P$ is the iteration times, $k$ is type of data partitions, and $L$ is assumption coefficient.

***Proof:*** The details can be found in our work which has been uploaded in our lab server.

### 4.2 Two-stage dynamic coded strategy

In this section, we will show our two-stage dynamic coded strategy for maximizing arrival data in detail. Firstly, our objective **P1** is to maximize the arrival data based on the past straggler behavior, code words assignment and data assignment. We specify how to design the coded strategy with the consideration of the heterogeneity worker. In the first phase, $M_1$ workers start computing the partial gradients from the data partitions and $K_c$ out of $K$ gradients which were computed by $K_c$ out of $M_1$ workers in the end of the first phase. In the second phase, $M_1 - M_c$ workers continue their gradients computation which were started in the first phase and $M - M_1$ workers start computing the partial gradients in the second phase depending upon which $K - K_c$ gradients were computed. Conventionally coded strategy should start in the first phase, in our two stage dynamic coded strategy we start in the second phase. Specifically, we random select $M_1$ workers in the first phase. After finishing executing the gradients of each workers for the duration time $T_{comp,<i>}$ in each epoch, it will send the partial result to the master, $M_1 - M_c$ worker continue their gradients computation and $M - M_1$ use coding strategy in the second phase if $K_c < K$. In other word, coding strategy is not triggered if $K_c = K$, since we are robust with stragglers in the first phase and complete computations for all gradient partitions.

Take Fig. 4 for an example, we illustrate the procedure of TSDCFL with 3 workers and i server. For instance, we start worker 1 and worker2, after worker 1 send the parameter to server within duration time in first stage. In second stage, worker 2 continue to update the parameters and woker3 are coded with worker 2 from remaining data partitions g2, until two worker send local update parameter, the epoch1 is finished. In epoch3，the worker2 and worker3 send the local parameter to the server in the first stage, we can say that we are robust to the stragglers in the first stage and encoding scheme is not triggered in second stage, which improve the efficiency and utilization of the workers.

In order to tolerate $s_{<i>}$ straggler in each epoch, each of $K - K_c$ gradients partitions to be assigned to at least $s_{<i>} + 1$ workers. There are total $(K - K_c)(s_{<i>} + 1) - \sum_{l=1}^{M_1 - M_c} n_l$ copies of data partitions, here $n_l$ is the proportions to the worker in the first phase. We now assume that the performance of workers are heterogeneous. We define the $W_m$ as the number of tasks finished per unit of time, and then define $w_m$ by normalizing $W_m$, i.e., $w_m = \frac{W_m}{\sum_{l=1}^{M-M_1} W_l}$. We define the $n_m$ as the proportions to the worker in the second phase, hence, we have

$$n_m = ((K - K_c)(s_{<i>} + 1) - \sum_{l=1}^{M_1 - M_c} n_l) \frac{W_m}{\sum_{l=1}^{M-M_1} W_l} \quad (16)$$

Once the $M_1$ is fixed, we can assign remaining $(K - K_c)(s_{<i>} + 1) - \sum_{l=1}^{M_1 - M_c} n_l$ to the workers in a second phase. The support structure of coding matrix is

$$\text{supp}(B_{(M-M_c) \times (K-K_c)}) = [b_1, b_2, ..., b_{M-M_c}] \quad (17)$$

Example 1. we consider a 5-workers system with one straggler, we construct the $B$ as:

$$\text{supp}(B_{5 \times 7}) = \begin{bmatrix} * & * & * & * & 0 & 0 & 0 \\ 0 & * & * & 0 & 0 & 0 & 0 \\ 0 & 0 & 0 & * & * & * & 0 \\ 0 & 0 & 0 & 0 & \bullet & 0 & \bullet \\ 0 & 0 & 0 & 0 & 0 & \bullet & \bullet \end{bmatrix}$$

Where $*$ denotes the data partitions assigned to worker $[m_1, m_2, m_3]$ in the first phase and $\bullet$ denotes the data partitions assigned to worker $[m_4, m_5]$ in the second phase. We assume that worker $m_1$ complete the gradient computation $[g_1, g_2, g_3, g_4]$ in the first phase and there are $[g_5, g_6, g_7]$ data partitions coded in the second phase. By reducing the rows and columns of the matrix, we can get the new coding matrix from above matrix

$$\text{supp}(B_{4 \times 3}) = \begin{bmatrix} 0 & 0 & 0 \\ * & * & 0 \\ \bullet & 0 & \bullet \\ 0 & \bullet & \bullet \end{bmatrix}$$

Given the structure of B, we need to assign code words to worker and introduce how to construct B such it can satisfy the span condition. We divide matrix B into two parts corresponding to an auxiliary matrix $A \in R^{(s_{<i>}+1) \times (M-M_c)}$ and an auxiliary vector $C \in R^{(1 \times M_c)}$, which follow the properties:

**T1**: for any the $s_{<i>} + 1$ columns of A is linearly independent. This property ensures that each type of data partitions is done by at least $s_{<i>} + 1$ workers.

**T2**: there exists a vector $D \in R^{(1) \times (s_{<i>}+1)}$ that guarantees that $s_{<i>}$ positions can be eliminated and 1 position remains. We then have that each data partitions belongs to non-stragglers.

**T3**: there exists a vector $C \in R^{(1 \times M_c)}$ that guarantees $CB^{M_c \times K_c} = 1^{1 \times K_c}$.

**Lemma 2.** *for a matrix $A \in R^{(s_{<i>}+1)\times(M-M_c)}$ have property* (T1), *there exists a vector $D \in R^{(1)\times(s_{<i>}+1)}$ have property* (T2) *and there exists a vector $C \in R^{(1\times M_c)}$ have property* (T3) *with the support structure of coding matrix satisfies span condition.*

***Proof:*** Our proof follows the following steps. First we randomly select a part of workers for gradient computation. When the deadline is reached, some workers $M_c$ finish the gradient computation and $K_c$ gradients calculation is completed at this moment. Since there exists a vector $C \in R^{(1\times M_c)}$ that guarantees $CB^{M_c \times K_c} = 1^{1\times K_c}$, we have $B^{M_c \times K_c}$ which satisfies span condition. Then we have $B_{(M-M_c)\times(K-K_c)}$ which satisfies span condition. For remaining gradient $k = 1,2...,K-K_c$ corresponding to each column of $B_{(M-M_c)\times(K-K_c)}$, since any the $s_{<i>}+1$ columns of A is linearly independent, we can get A is non-singular. We can get the code words of each column: $AB_{(M-M_c)\times(K-K_c)} = 1^{(s_{<i>}+1)(K-K_c)}$. Next we show that the $B_{(M-M_c)\times(K-K_c)}$ satisfies the span condition. Since there exists a vector $D \in R^{(1)\times(s_{<i>}+1)}$, hence $DAB_{(M-M_c)\times(K-K_c)} = 1^{1\times(K-K_c)}$. Therefore, B satisfies span condition. The proof is completed.

### 4.3 Fairness transmission strategy

Given the arrival data in our system, the next objective is to maximize the throughput and fairness in our coding system, to achieve which reduce the likelihood number of stragglers in communication phase. Specifically, we aim to discourage individual workers to consume excessive energy for little utility gain. It is challenging to get optimal values due to the following reasons. First, the arrival data of gradient of workers are time-varying and unknown a priori, hence it is hard to make decisions to admission data. Secondly, the backlog of gradient data is instability in the long run. In this section, we resort to Lyapunov optimization to solve the optimization problem.

**Lemma 3.** *The optimization problem* **P2** *is equivalent to*

$$\text{P3: max } \overline{\log(1+y)}$$
$$\text{s.t. C1-C5 and C6: } \overline{y_m} < \overline{d_{m,<i>}} \quad (18)$$

***Proof***: the problem is solved by constructing non-negative auxiliary variables corresponding to admission data. We need to prove that maximum **P3** is equivalent to the **P2**. Since **P2** is concave function, this function is monotonically increasing function, **C6** guarantees $\log(1+\overline{d}^*) \geq \log(1+\overline{y})$, by applying Jensen's inequality on the concave function, we can get the $\log(1+\overline{y}) \geq \overline{\log(1+y)}$ and we can conclude that **P3** is equivalent to the **P2**.

Next, we construct a virtual queue for admission data, which can be expressed as

$$H_{m,<i>}(t+1) = \max[H_{m,<i>}(t) + y_{m,<i>} - d_{m,<i>}, 0]()$$

This virtual queue is stable if and only if $H_{m,<i>}(t)$ can transform into a queue stability problem. Then, this queue coupled with other queue can be addressed with stability constrains. In specific, given the queue dynamic, we can define a perturbed Lyapunov function as

$$L(t) = \frac{1}{2}\left\{\sum_{m \in M_{non\_stragglers}}\left[H_{m,<i>}(t)^2 + Q_{m,<i>}(t)^2 + E_{m,<i>}(t)^2 + R_{m,<i>}(t)^2\right] + R_{server,<i>}(t)^2\right\}$$
(19)

The above function is always non-negative and $L(t) = 0$ if and only if queue size equal 0. Let

$$\Theta(t) = (\{H_{m,<i>}(t), Q_{m,<i>}(t), E_{m,<i>}(t), R_{m,<i>}(t), \forall m \in M_{non\_stragglers}\}, R_{server,<i>}(t))$$

indicate the instantaneous backlogs size at time slot. Then, we can define the one-slot drift-plus-penalty function as

$$\Delta_V(t) = \mathbf{E}\{L(t+1) - L(t)|\Theta(t)\} - V\mathbf{E}\left\{\sum_{M_{non\_stragglers}} \log(1+y_m(t)|\Theta(t)\right\}$$

A smaller value $\Delta_V(t)$ indicates stable system utility and backlogs. We aim to minimize upper-bound of $\Delta_V(t)$ so as to prevent the growth of backlogs and increase stability of system.

**Lemma 4.** *The optimization problem $\Delta_V(t)$ is upper bounded by*

$$\Delta_V(t) \leq \frac{1}{2}\sum_{m \in M_{non\_stragglers}}(3D_{m,<i>}^{\max} + (r_{m,<i>}^{\max}T)^2 + ((p_mT)^2 + (f_m^{\max}\delta_m)^2 + (E_{\max,<i>}^H)^2 + (f_m^{\max})^2)$$

$$+ \frac{1}{2}((F^{\max})^2 + \sum_{m \in M_{non\_stragglers}}(r_{m,<i>}^{\max}T\xi_m)^2) - V\mathbf{E}\left\{\sum_{m \in M_{non\_stragglers}}\log(1+y_m(t)|\Theta(t)\right\}$$

$$+ \sum_{m \in M_{non\_stragglers}}Q_{m,<i>}(t)\mathbf{E}\{(d_{m,<i>}(t) - c_{m,<i>}(t)|\Theta(t)\} + \sum_{m \in M_{non\_stragglers}}H_{m,<i>}(t)\mathbf{E}\{(y_{m,<i>}(t) - d_{m,<i>}(t))|\Theta(t)\} \quad (20)$$

$$+ \sum_{m \in M_{non\_stragglers}}E_{m,<i>}(t)\mathbf{E}\{(e_{m\_store,<i>}(t) - (e_{m\_up,<i>}(t) + e_{m\_com,<i>}(t)))|\Theta(t)\}$$

$$- \sum_{m \in M_{non\_stragglers}}R_{m,<i>}(t)\mathbf{E}\{f_m(t)|\Theta(t)\} + R_{server,<i>}(t)\mathbf{E}\{(\sum_{m \in M_{non\_stragglers}}c_{m,<i>}(t)\xi_m - F(t))|\Theta(t)\}$$

***Proof:*** The details can be found in our work which has been uploaded in our lab server.

The upper bound allows us to make decisions by minimizing the value on upper bound. Therefore, we can maximize the throughput and fairness in our coding scheme, to achieve, which can reduce the likelihood number of stragglers in communication phase. Therefore, by exploiting **Lemma 4**, it is easy to divide **P3** into independent terms, in which can be decomposed equivalently into the following sub-problems (**P4**, **P5**, **P6** and **P7**)

$$\begin{cases} \text{P4:} & \max V \left\{ \sum_{m \in M_{non\_stragglers}} \log(1 + y_m(t)) \right\} - \sum_{m \in M_{non\_stragglers}} H_{m,<i>}(t) \{(y_{m,<i>}(t) - d_{m,<i>}(t))\} \\ \text{P5:} & \min \sum_{m \in M_{non\_stragglers}} (Q_{m,<i>}(t) - H_{m,<i>}(t)) d_{m,<i>}(t) \\ \text{P6:} & \min \sum_{m \in M_{non\_stragglers}} E_{m,<i>}(t)(e_{m\_store,<i>}(t) - (e_{m\_up,<i>}(t) + e_{m\_com,<i>}(t))) \\ \text{P7:} & \max \sum_{m \in M_{non\_stragglers}} E_{m,<i>}(t) e_{m\_up,<i>}(t) + (Q_{m,<i>}(t) - R_{m,<i>}(t)(t) \xi_m) c_{m,<i>} \end{cases}$$

The parameters of the drift-plus-penalty upper bound can be viewed as an upper bound of $\Delta_V(t)$, we are able to maximize the throughput and sum of the system drift by making independent and sequential decision on $\upsilon_{m,<i>}(t)$, $d_m(t)$, $y_m(t)$, and $e_{m\_store,<i>}$. Therefore, by setting value of these parameters, we can optimize the four separated conditional terms on **P4**, **P5**, **P6** and **P7**. By tuning the value, all the queue states are optimized in each time slot. Next, we address these four sub-problems by the auxiliary variable determination, optimal schedules for admission data, admission time and scheduling, optimal energy schedule.

1) **Auxiliary variable determination**: given the condition $y_m < D_{m,<i>}(t)$ and observed queue state, it turn to maximize

$$V \left\{ \sum_{m \in M_{non\_stragglers}} \log(1 + y_m(t)) \right\} - \sum_{m \in M_{non\_stragglers}} H_{m,<i>}(t)(y_{m,<i>}(t)$$

the optimization is able to solve by setting the $y_m(t)$. The first order of **P4** is given by $\frac{V}{(1+y_m(t))\ln 2} - H_{m,<i>}(t)$. It is readily solved since the optimal is either at the stationary point. If $\frac{V}{\ln 2} - H_{m,<i>}(t) \leq 0$, the **P4** is closed form functions and we can get $y_m^*(t) = 0$. Otherwise, the **P4** increases monotonically and we have $y_m^*(t) = \min(\frac{V}{H_{m,<i>}(t)\ln 2} - \frac{1}{\ln 2}, D_{m,<i>}(t))$.

2) **Optimal schedules for admission data:** Given the queuing states $Q_{m,<i>}(t)$ and $H_{m,<i>}(t)$, it is easy to determine a set of $d_m(t)$ to maximize $(Q_{m,<i>}(t) - H_{m,<i>}(t))d_{m,<i>}(t)$ within the feasible interval of $d_m(t)$. Specially, for the $m^{th}$ worker, a simple scheduling strategy is to send all the admitted data. If $Q_{m,<i>}(t) \geq H_{m,<i>}(t)$ and $d_m^*(t) = 0$. Otherwise, $d_m^*(t) = D_{m,<i>}(t)$.

3) **Admission time and scheduling:** Considering the coupling constraint $\sum_{m \in M_{non\_stragglers,<i>}} \upsilon_{m,<i>}(t) \leq TL(t)$, it is difficult to determine the proper admission time. While the number of sub-channels is far less than the workers, we can consider this problem as knapsack problem. Specifically, we determine the admission time sequentially based on the admission data, and admission time sequentially as long as the constraint is satisfied. Admission time should not transmit more than harvested energy $E_{m,<i>}(t)/p_m$ and data backlog $Q_{m,<i>}(t)/r_m(t)$. Note that, queues with negative is not considered, and corresponding $\upsilon_{m,<i>}(t) = 0$.

4) **Optimal energy schedule:** Given the queue state, it is easy to obtain optimal energy intake by minimizing $E_{m,<i>}(t)(e_{m\_store,<i>}(t) - (e_{m\_up,<i>}(t) + e_{m\_com,<i>}(t)))$ based on instantaneous queue size, i.e., obtaining the feasible interval $e_{m\_store,<i>}(t) \in [0, E_{m,<i>}(t)]$.

## IV. PERFORMANCE EVALUATION

In the experiment, we try to use the KubeEdge with connecting multiple machine, including one cloud host and a number of edge node. KubeEdge platform is an edge computing framework base on Kubernetes, which provides management compute resource, service deployment, and runtime scheduling and operation capabilities. We build a 7-node consisting of 1 parameter server and 6 workers. All of nodes are established on KubeEdge. We mainly use different CPU cores to simulate the training parallelism. To simulate the stragglers, we inject one or two stragglers into training per epoch. All algorithms are implemented on Python 3.5 and PyTorch 0.4. Furthermore, we rely on an additional library cjl-test implemented on PyTorch to partition the dataset and obtain gradients. All algorithm steps run on the CPU, except for loading training data from disk and communicating gradients through the network. To compare the performance of different methods, experiments were conducted on 6 nodes. We compare our algorithms to **Cyclic Repetition Scheme** and **Fractional Repetition Scheme.** We use image classification datasets including Mnist and Cifar10 as our benchmarks. We set the batch size as 128 and initial learning rate is 0.01. The entire training process is restricted in 10000 iterations. In this paper, we mainly adopt time-based efficiency and epoch based efficiency as the metrics. Time-based efficiency is affected by the arrival data and admission data in our system. We adopt two stage dynamic coding scheme and fairness transmission scheme to improve efficiency.

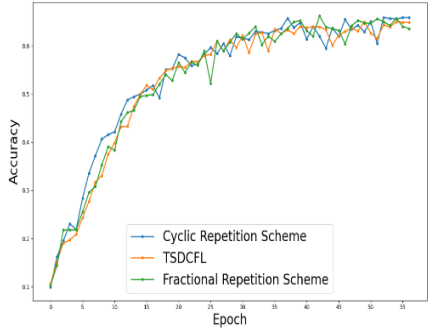

Fig. 5 a- Accuracy with epoch.

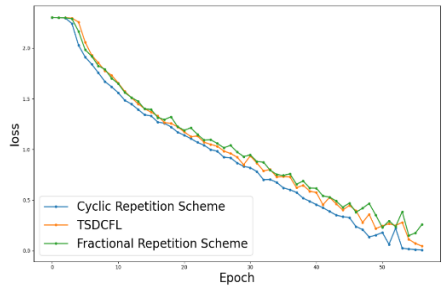

Fig. 5 b-Loss with epoch.

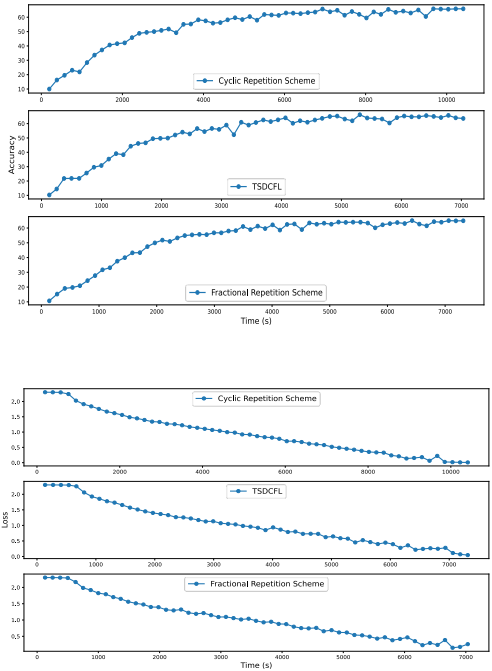

Fig. 5 c- Accuracy and Fig. 5 d-loss with time.

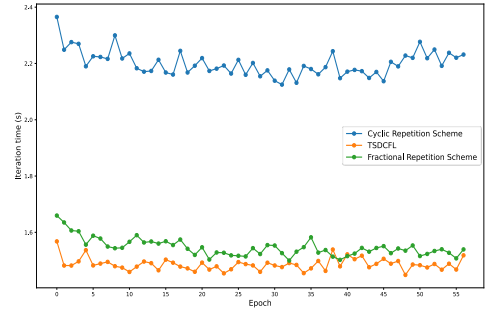

Fig. 5 e- Iteration time with epoch.
Fig. 5 Accuracy, loss, and iteration time over TSDCFL in Cifar10.

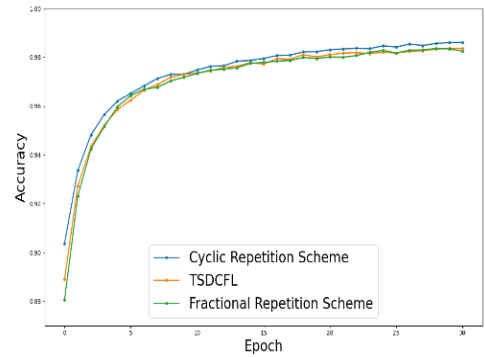

Fig. 6 a- Accuracy with epoch.

As shown in Fig. 5 and Fig. 6, we can see that **TSDCFL** is the same epoch based efficiency as the algorithms. The main reason is that all algorithms adopt the full data partitions and synchronous scheme in each iteration, as the consequence, epoch Based have the same converge rate. After compared with other methods with epoch based efficiency, we conduct the experiments on 6 nodes with (2 cores, 2cores, 4 cores, 4 cores, 8 cores, 8 cores) for time-based efficiency of two models. The results show that TSDCFL preforms the better than other algorithms due to its tolerance to the stragglers, higher utilization of the computation resource and fairness transmission among workers. Specifically, as shown in Fig. 5e and Fig. 6e, the iteration time of TSDCFL is lower than the other algorithms. To be summarized, the algorithm with TSDCFL achieves the performance improvement by reducing the computation time and transmission time while keeping the convergence rate.

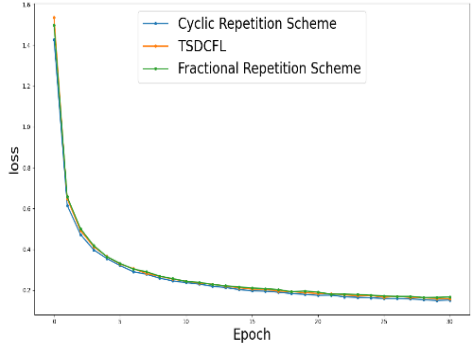

Fig. 6 b-Loss with epoch.

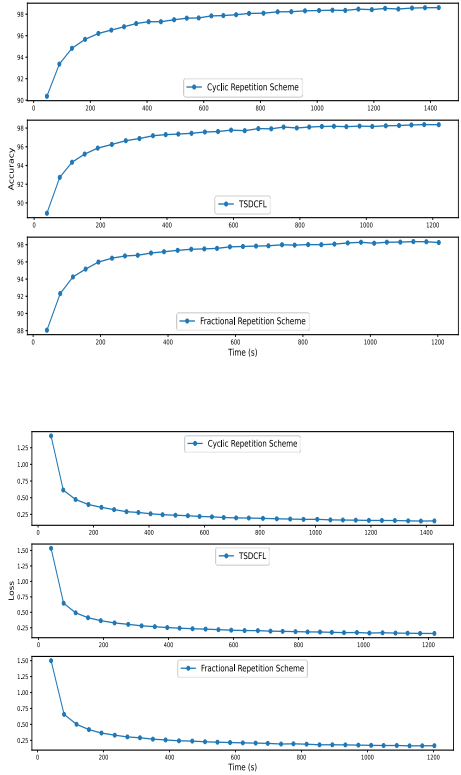

Fig. 6 c- Accuracy and Fig. 6 d-loss with time.

## V. CONCLUSION

In this paper, we propose a novel coded FEL to mitigate stragglers for synchronous gradient with a two-stage dynamic scheme to enhance accuracy and resource utilization in the FEL system, where we start with part of workers for a duration of before starting the second stage, and on completion of at the first stage, we start remaining workers in the second stage. In the proposed scheme, the computation latency and transmission latency is essential and quantitatively analyzed. Then the dynamically coded coefficients scheme is proposed which is based on historical information including worker completion time. For performance optimization of FEL, a Lyapunov function is designed to maximize admission data balancing fairness and two stage dynamic coding scheme is designed to maximize arrival data among workers. The extensive experiments demonstrate the effectiveness of the proposed scheme, which can achieve the performance improvement by reducing the time and transmission time while keeping the convergence rate.

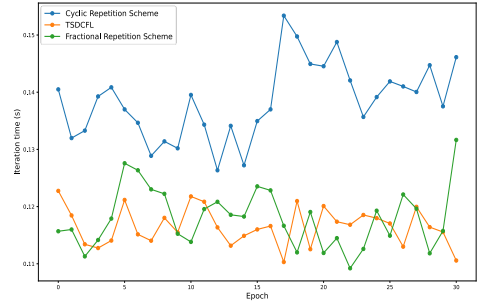

Fig. 6 e- Iteration time with epoch.
Fig. 6 Accuracy, loss, and iteration time over TSDCFL in Mnist.

$$\mathrm{E}\left[F(w_{<i>})\right]-F(w_{<i-1>})$$

$$\leq \langle g(w_{<i-1>}), w_{<i>}-w_{<i-1>}\rangle + \frac{L}{2}\|w_{<i>}-w_{<i-1>}\|^2$$

$$=-\mathrm{E}\left[\bar{\eta}\langle g(w_{<i-1>}), \frac{1}{K}(\sum_{k=1}^{K_1}\sum_{m=1}^{M_1}a_{m,<i-1>}b_{m,k}g_k(i-1-\tau_{m,<i-1>})+\sum_{K=1}^{K_2}\sum_{m=1}^{M_2}a_{m,<i>}b_{m,k}g_k(i-1-\tau_{m,<i-1>}))\rangle\right]$$

$$+\frac{L\bar{\eta}^2}{2}\mathrm{E}\left[\left\|\frac{1}{K}(\sum_{k=1}^{K_1}\sum_{m=1}^{M_1}a_{m,<i-1>}b_{m,k}g_k(i-1-\tau_{m,<i-1>})+\sum_{K=1}^{K_2}\sum_{m=1}^{M_2}a_{m,<i-1>}b_{m,k}g_k(i-1-\tau_{m,<i-1>}))\right\|^2\right]$$

$$=-\bar{\eta}\langle g(w_{<i-1>}), \frac{1}{K}(\sum_{k=1}^{K_1}\sum_{m=1}^{M_1}a_{m,<i-1>}b_{m,k}g_k(i-1-\tau_{m,<i-1>})+\sum_{K=1}^{K_2}\sum_{m=1}^{M_2}a_{m,<i>}b_{m,k}g_k(i-1-\tau_{m,<i-1>}))\rangle$$

$$+\frac{L\bar{\eta}^2}{2}\mathrm{E}\left[\left\|\frac{1}{K}(\sum_{k=1}^{K_1}\sum_{m=1}^{M_1}a_{m,<i-1>}b_{m,k}g_k(i-1-\tau_{m,<i-1>})+\sum_{K=1}^{K_2}\sum_{m=1}^{M_2}a_{m,<i-1>}b_{m,k}g_k(i-1-\tau_{m,<i-1>}))\right\|^2\right]$$

$$=-\frac{1}{2}\bar{\eta}(\|g(w_{<i-1>})\|^2+\left\|\frac{1}{K}(\sum_{k=1}^{K_1}\sum_{m=1}^{M_1}a_{m,<i-1>}b_{m,k}g_k(i-1-\tau_{m,<i-1>})+\sum_{k=1}^{K_2}\sum_{m=1}^{M_2}a_{m,<i-1>}b_{m,k}g_k(i-1-\tau_{m,<i-1>}))\right\|^2$$

$$-\left\|g(w_{<i-1>})-\frac{1}{K}(\sum_{k=1}^{K_1}\sum_{m=1}^{M_1}a_{m,<i-1>}b_{m,k}g_k(i-1-\tau_{m,<i-1>})+\sum_{k=1}^{K_2}\sum_{m=1}^{M_2}a_{m,<i-1>}b_{m,k}g_k(i-1-\tau_{m,<i-1>}))\right\|^2)$$

$$+\frac{L\bar{\eta}^2}{2}\mathrm{E}\left[\left\|\frac{1}{K}(\sum_{k=1}^{K_1}\sum_{m=1}^{M_1}a_{m,<i-1>}b_{m,k}g_k(i-1-\tau_{m,<i-1>})+\sum_{k=1}^{K_2}\sum_{m=1}^{M_2}a_{m,<i-1>}b_{m,k}g_k(i-1-\tau_{m,<i-1>}))\right\|^2\right]$$

$$=-\frac{1}{2}\bar{\eta}(\|g(w_{<i-1>})\|^2+\left\|\frac{1}{K}(\sum_{k=1}^{K_1}\sum_{m=1}^{M_1}a_{m,<i>}b_{m,k}g_k(i-1-\tau_{m,<i-1>})+\sum_{k=1}^{K_2}\sum_{m=1}^{M_2}a_{m,<i-1>}b_{m,k}g_k(i-1-\tau_{m,<i-1>}))\right\|^2$$

$$-\left\|g(w_{<i-1>})-\frac{1}{K}(\sum_{k=1}^{K_1}\sum_{m=1}^{M_1}a_{m,<i-1>}b_{m,k}g_k(i-1-\tau_{m,<i-1>})+\sum_{k=1}^{K_2}\sum_{m=1}^{M_2}a_{m,<i-1>}b_{m,k}g_k(i-1-\tau_{m,<i-1>}))\right\|^2)$$

$$+\frac{L\bar{\eta}^2}{2}\mathrm{E}\left[\left\|\frac{1}{K}(\sum_{k=1}^{K_1}\sum_{m=1}^{M_1}a_{m,<i-1>}b_{m,k}(g_k(i-1-\tau_{m,<i-1>})-g(i-1-\tau_{m,<i-1>}))\right.\right.$$

$$\left.\left.+\sum_{k=1}^{K_2}\sum_{m=1}^{M_2}a_{m,<i-1>}b_{m,k}(g_k(i-1-\tau_{m,<i-1>})-g(i-1-\tau_{m,<i-1>})))+\frac{1}{K}\sum_{k=1}^{K}\sum_{m=1}^{M}a_{m,<i-1>}b_{m,k}g(i-1-\tau_{m,<i-1>})\right\|^2\right]$$

$$\leq -\frac{1}{2}\bar{\eta}(\|g(w_{<i-1>})\|^2+\left\|\frac{1}{K}(\sum_{k=1}^{K_1}\sum_{m=1}^{M_1}a_{m,<i>}b_{m,k}g_k(i-1-\tau_{m,<i-1>})+\sum_{k=1}^{K_2}\sum_{m=1}^{M_2}a_{m,<i-1>}b_{m,k}g_k(i-1-\tau_{m,<i-1>}))\right\|^2$$

$$-\left\|g(w_{<i-1>})-\frac{1}{K}(\sum_{k=1}^{K_1}\sum_{m=1}^{M_1}a_{m,<i-1>}b_{m,k}g_k(i-1-\tau_{m,<i-1>})+\sum_{k=1}^{K_2}\sum_{m=1}^{M_2}a_{m,<i-1>}b_{m,k}g_k(i-1-\tau_{m,<i-1>}))\right\|^2)$$

$$+\frac{3L\bar{\eta}^2}{2}\mathrm{E}\left[\left\|\frac{1}{K}(\sum_{k=1}^{K_1}\sum_{m=1}^{M_1}a_{m,<i-1>}b_{m,k}(g_k(i-1-\tau_{m,<i-1>})-g(i-1-\tau_{m,<i-1>}))\right.\right.$$

$$\left.\left.+\sum_{k=1}^{K_2}\sum_{m=1}^{M_2}a_{m,<i-1>}b_{m,k}(g_k(i-1-\tau_{m,<i-1>})-g(i-1-\tau_{m,<i-1>}))\right\|^2+\left\|\frac{1}{K}\sum_{k=1}^{K}\sum_{m=1}^{M}g(i-1-\tau_{m,<i-1>})\right\|^2\right]$$

$$\leq -\frac{1}{2}\bar{\eta}(\|g(w_{<i-1>})\|^2+\left\|\frac{1}{K}(\sum_{k=1}^{K_1}\sum_{m=1}^{M_1}a_{m,<i-1>}b_{m,k}g_k(i-1-\tau_{m,<i-1>})+\sum_{k=1}^{K_2}\sum_{m=1}^{M_2}a_{m,<i-1>}b_{m,k}g_k(i-1-\tau_{m,<i-1>}))\right\|^2$$

$$-\left\|g(w_{<i-1>})-\frac{1}{K}(\sum_{k=1}^{K_1}\sum_{m=1}^{M_1}a_{m,<i-1>}b_{m,k}g_k(i-1-\tau_{m,<i-1>})+\sum_{k=1}^{K_2}\sum_{m=1}^{M_2}a_{m,<i-1>}b_{m,k}g_k(i-1-\tau_{m,<i-1>}))\right\|^2)$$

$$+\frac{3L\bar{\eta}^2}{2K^2}\mathrm{E}\left[(\sum_{k=1}^{K_1}\sum_{m=1}^{M_1}a_{m,<i-1>}^2 b_{m,k}^2 \left\|(g_k(i-1-\tau_{m,<i-1>})-g(i-1-\tau_{m,<i-1>}))\right\|^2\right.$$

$$\left.+\sum_{k=1}^{K_2}\sum_{m=1}^{M_2}a_{m,<i-1>}^2 b_{m,k}^2 \left\|(g_k(i-1-\tau_{m,<i-1>})-g(i-1-\tau_{m,<i-1>}))\right\|^2)+\left\|\frac{1}{K}\sum_{k=1}^{K}\sum_{m=1}^{M}g(i-1-\tau_{m,<i-1>})\right\|^2\right]$$

$$\leq -\frac{1}{2}\bar{\eta}(\left\|g(w_{<i-1>})\right\|^2+\left\|\frac{1}{K}(\sum_{k=1}^{K_1}\sum_{m=1}^{M_1}a_{m,<i-1>}b_{m,k}g_k(i-1-\tau_{m,<i-1>})+\sum_{k=1}^{K_2}\sum_{m=1}^{M_2}a_{m,<i-1>}b_{m,k}g_k(i-1-\tau_{m,<i-1>}))\right\|^2$$

$$-\left\|g(w_{<i-1>})-\frac{1}{K}(\sum_{k=1}^{K_1}\sum_{m=1}^{M_1}a_{m,<i-1>}b_{m,k}g_k(i-1-\tau_{m,<i-1>})+\sum_{k=1}^{K_2}\sum_{m=1}^{M_2}a_{m,<i-1>}b_{m,k}g_k(i-1-\tau_{m,<i-1>}))\right\|^2)$$

$$+\frac{3L\bar{\eta}^2}{2K^2}(\sum_{k=1}^{K_1}\sum_{m=1}^{M_1}a_{m,<i-1>}^2 b_{m,k}^2\varsigma^2+\sum_{k=1}^{K_2}\sum_{m=1}^{M_2}a_{m,<i-1>}^2 b_{m,k}^2\varsigma^2)+\frac{3L\bar{\eta}^2}{2}\left\|\frac{1}{K}\sum_{k=1}^{K}\sum_{m=1}^{M}g(i-1-\tau_{m,<i-1>})\right\|^2$$

$$= -\frac{1}{2}\bar{\eta}(\left\|g(w_{<i-1>})\right\|^2+\left\|\frac{1}{K}(\sum_{k=1}^{K_1}\sum_{m=1}^{M_1}a_{m,<i-1>}b_{m,k}g_k(i-1-\tau_{m,<i-1>})+\sum_{k=1}^{K_2}\sum_{m=1}^{M_2}a_{m,<i-1>}b_{m,k}g_k(i-1-\tau_{m,<i-1>}))\right\|^2$$

$$-\left\|\frac{1}{K}(\sum_{k=1}^{K_1}\sum_{m=1}^{M_1}a_{m,<i-1>}b_{m,k}(g(w_{<i-1>})-g_k(i-1-\tau_{m,<i-1>}))+\sum_{k=1}^{K_2}\sum_{m=1}^{M_2}a_{m,<i-1>}b_{m,k}(g(w_{<i-1>})-g_k(i-1-\tau_{m,<i-1>})))\right\|^2)$$

$$+\frac{3L\bar{\eta}^2}{2K^2}(\sum_{k=1}^{K_1}\sum_{m=1}^{M_1}a_{m,<i-1>}^2 b_{m,k}^2\varsigma^2+\sum_{k=1}^{K_2}\sum_{m=1}^{M_2}a_{m,<i-1>}^2 b_{m,k}^2\varsigma^2)+\frac{3L\bar{\eta}^2}{2}\left\|\frac{1}{K}\sum_{k=1}^{K}\sum_{m=1}^{M}g(i-1-\tau_{m,<i-1>})\right\|^2$$

$$\leq -\frac{1}{2}\bar{\eta}(\left\|g(w_{<i-1>})\right\|^2+\left\|\frac{1}{K}(\sum_{k=1}^{K_1}\sum_{m=1}^{M_1}a_{m,<i-1>}b_{m,k}g_k(i-1-\tau_{m,<i-1>})+\sum_{k=1}^{K_2}\sum_{m=1}^{M_2}a_{m,<i-1>}b_{m,k}g_k(i-1-\tau_{m,<i-1>}))\right\|^2$$

$$-\frac{1}{K}(\sum_{k=1}^{K_1}\sum_{m=1}^{M_1}a_{m,<i-1>}b_{m,k}\left\|(g(w_{<i-1>})-g_k(i-1-\tau_{m,<i-1>}))\right\|^2+\sum_{k=1}^{K_2}\sum_{m=1}^{M_2}a_{m,<i-1>}b_{m,k}\left\|(g(w_{<i-1>})-g_k(i-1-\tau_{m,<i-1>}))\right\|^2)$$

$$+\frac{3L\bar{\eta}^2}{2K^2}(\sum_{k=1}^{K_1}\sum_{m=1}^{M_1}a_{m,<i-1>}^2 b_{m,k}^2\varsigma^2+\sum_{k=1}^{K_2}\sum_{m=1}^{M_2}a_{m,<i-1>}^2 b_{m,k}^2\varsigma^2)+\frac{3L\bar{\eta}^2}{2}\left\|\frac{1}{K}\sum_{k=1}^{K}\sum_{m=1}^{M}g(i-1-\tau_{m,<i-1>})\right\|^2$$

$$\leq -\frac{1}{2}\bar{\eta}(\left\|g(w_{<i-1>})\right\|^2+\left\|\frac{1}{K}(\sum_{k=1}^{K_1}\sum_{m=1}^{M_1}a_{m,<i-1>}b_{m,k}g_k(i-1-\tau_{m,<i-1>})+\sum_{k=1}^{K_2}\sum_{m=1}^{M_2}a_{m,<i-1>}b_{m,k}g_k(i-1-\tau_{m,<i-1>}))\right\|^2$$

$$-\frac{1}{K}L^2(\sum_{k=1}^{K_1}\sum_{m=1}^{M_1}a_{m,<i-1>}b_{m,k}\left\|(w_{<i-1>}-w_{<i-1-\tau_{m,<i-1>}>})\right\|^2+\sum_{k=1}^{K_2}\sum_{m=1}^{M_2}a_{m,<i-1>}b_{m,k}\left\|((w_{<i-1>}-w_{<i-1-\tau_{m,<i-1>}>}))\right\|^2)$$

$$+\frac{L\bar{\eta}^2}{K^2}(\sum_{k=1}^{K_1}\sum_{m=1}^{M_1}a_{m,<i-1>}^2 b_{m,k}^2\varsigma^2+\sum_{k=1}^{K_2}\sum_{m=1}^{M_2}a_{m,<i-1>}^2 b_{m,k}^2\varsigma^2)+L\bar{\eta}^2\left\|\frac{1}{K}\sum_{k=1}^{K}\sum_{m=1}^{M}g(i-1-\tau_{m,<i-1>})\right\|^2$$

$$\leq -\frac{1}{2}\bar{\eta}(\left\|g(w_{<i-1>})\right\|^2+\left\|\frac{1}{K}(\sum_{k=1}^{K_1}\sum_{m=1}^{M_1}a_{m,<i-1>}b_{m,k}g_k(i-1-\tau_{m,<i-1>})+\sum_{k=1}^{K_2}\sum_{m=1}^{M_2}a_{m,<i-1>}b_{m,k}g_k(i-1-\tau_{m,<i-1>}))\right\|^2$$

$$-\frac{1}{K}L^2(\sum_{k=1}^{K_1}\sum_{m=1}^{M_1}a_{m,<i-1>}b_{m,k}\left\|(w_{<i-1>}-w_{<i-1-\tau_{\max}>})\right\|^2+\sum_{k=1}^{K_2}\sum_{m=1}^{M_2}a_{m,<i-1>}b_{m,k}\left\|((w_{<i-1>}-w_{<i-1-\tau_{\max}>}))\right\|^2)$$

$$+\frac{3L\bar{\eta}^2}{2K^2}(\sum_{k=1}^{K_1}\sum_{m=1}^{M_1}a_{m,<i-1>}^2 b_{m,k}^2\varsigma^2+\sum_{k=1}^{K_2}\sum_{m=1}^{M_2}a_{m,<i-1>}^2 b_{m,k}^2\varsigma^2)+\frac{3L\bar{\eta}^2}{2}\left\|\frac{1}{K}\sum_{k=1}^{K}\sum_{m=1}^{M}g(i-1-\tau_{m,<i-1>})\right\|^2$$

$$= -\frac{1}{2}\bar{\eta}(\left\|g(w_{<i-1>})\right\|^2+\left\|\frac{1}{K}(\sum_{k=1}^{K_1}\sum_{m=1}^{M_1}a_{m,<i-1>}b_{m,k}g_k(i-1-\tau_{m,<i-1>})+\sum_{k=1}^{K_2}\sum_{m=1}^{M_2}a_{m,<i-1>}b_{m,k}g_k(i-1-\tau_{m,<i-1>}))\right\|^2$$

$$-L^2(\left\|(w_{<i-1>}-w_{<i-1-\tau_{\max}>})\right\|^2)+\frac{3L\bar{\eta}^2}{2K^2}(\sum_{k=1}^{K_1}\sum_{m=1}^{M_1}a_{m,<i-1>}^2 b_{m,k}^2\varsigma^2+\sum_{k=1}^{K_2}\sum_{m=1}^{M_2}a_{m,<i-1>}^2 b_{m,k}^2\varsigma^2)+\frac{3L\bar{\eta}^2}{2}\left\|\frac{1}{K}\sum_{k=1}^{K}\sum_{m=1}^{M}g(i-1-\tau_{m,<i-1>})\right\|^2$$

$$= -\frac{1}{2}\bar{\eta}(\|g(w_{<i-1>})\|^2 + \left\|\frac{1}{K}(\sum_{k=1}^{K_1}\sum_{m=1}^{M_1}a_{m,<i-1>}b_{m,k}g_k(i-1-\tau_{m,<i-1>}) + \sum_{k=1}^{K_2}\sum_{m=1}^{M_2}a_{m,<i-1>}b_{m,k}g_k(i-1-\tau_{m,<i-1>}))\right\|^2$$

$$-L^2(\left\|\sum_{n=i-1-\tau_{max}}^{n=i-1}(w_{<n>}-w_{<n-1>})\right\|^2) + \frac{3L\bar{\eta}^2}{2K^2}(\sum_{k=1}^{K_1}\sum_{m=1}^{M_1}a_{m,<i-1>}^2 b_{m,k}^2\varsigma^2 + \sum_{k=1}^{K_2}\sum_{m=1}^{M_2}a_{m,<i-1>}^2 b_{m,k}^2\varsigma^2)$$

$$+\frac{3L\bar{\eta}^2}{2}\left\|\frac{1}{K}\sum_{k=1}^{K}\sum_{m=1}^{M}g(i-1-\tau_{m,<i-1>})\right\|^2$$

$$= -\frac{1}{2}\bar{\eta}(\|g(w_{<i-1>})\|^2 + \left\|\frac{1}{K}(\sum_{k=1}^{K_1}\sum_{m=1}^{M_1}a_{m,<i-1>}b_{m,k}g_k(i-1-\tau_{m,<i-1>}) + \sum_{k=1}^{K_2}\sum_{m=1}^{M_2}a_{m,<i-1>}b_{m,k}g_k(i-1-\tau_{m,<i-1>}))\right\|^2$$

$$-L^2(\left\|\sum_{n=i-1-\tau_{max}}^{n=i-1}\frac{1}{K}\bar{\eta}(\sum_{k=1}^{K_1}\sum_{m=1}^{M_1}a_{m,<i-1>}b_{m,k}g_k(n-1-\tau_{m,<n-1>}) + \sum_{k=1}^{K_2}\sum_{m=1}^{M_2}a_{m,<i-1>}b_{m,k}g_k(n-1-\tau_{m,<n-1>})))\right\|^2)$$

$$+\frac{3L\bar{\eta}^2}{2K^2}(\sum_{k=1}^{K_1}\sum_{m=1}^{M_1}a_{m,<i-1>}^2 b_{m,k}^2\varsigma^2 + \sum_{k=1}^{K_2}\sum_{m=1}^{M_2}a_{m,<i-1>}^2 b_{m,k}^2\varsigma^2) + \frac{3L\bar{\eta}^2}{2}\left\|\frac{1}{K}\sum_{k=1}^{K}\sum_{m=1}^{M}g(i-1-\tau_{m,<i-1>})\right\|^2$$

$$= -\frac{1}{2}\bar{\eta}(\|g(w_{<i-1>})\|^2 + \left\|\frac{1}{K}(\sum_{k=1}^{K_1}\sum_{m=1}^{M_1}a_{m,<i-1>}b_{m,k}g_k(i-1-\tau_{m,<i-1>}) + \sum_{k=1}^{K_2}\sum_{m=1}^{M_2}a_{m,<i-1>}b_{m,k}g_k(i-1-\tau_{m,<i-1>}))\right\|^2$$

$$-L^2\left(\left\|\sum_{n=i-1-\tau_{max}}^{n=i-1}\frac{1}{K}\bar{\eta}(\sum_{k=1}^{K_1}\sum_{m=1}^{M_1}a_{m,<i-1>}b_{m,k}(g_k(n-1-\tau_{m,<n-1>}) - g(n-1-\tau_{m,<n-1>}))\right.\right.$$

$$\left.+\sum_{k=1}^{K_2}\sum_{m=1}^{M_2}a_{m,<i-1>}b_{m,k}(g_k(n-1-\tau_{m,<n-1>}) - g(n-1-\tau_{m,<n-1>}))) + \sum_{n=i-1-\tau_{max}}^{n=i-1}\frac{1}{K}\bar{\eta}\sum_{k=1}^{K}\sum_{m=1}^{M}a_{m,<i-1>}b_{m,k}g(n-1-\tau_{m,<n-1>})\right\|^2\right)$$

$$+\frac{3L\bar{\eta}^2}{2K^2}(\sum_{k=1}^{K_1}\sum_{m=1}^{M_1}a_{m,<i-1>}^2 b_{m,k}^2\varsigma^2 + \sum_{k=1}^{K_2}\sum_{m=1}^{M_2}a_{m,<i-1>}^2 b_{m,k}^2\varsigma^2) + \frac{3L\bar{\eta}^2}{2}\left\|\frac{1}{K}\sum_{k=1}^{K}\sum_{m=1}^{M}g(i-1-\tau_{m,<i-1>})\right\|^2$$

$$\leq -\frac{1}{2}\bar{\eta}(\|g(w_{<i-1>})\|^2 + \left\|\frac{1}{K}(\sum_{k=1}^{K_1}\sum_{m=1}^{M_1}a_{m,<i-1>}b_{m,k}g_k(i-1-\tau_{m,<i-1>}) + \sum_{k=1}^{K_2}\sum_{m=1}^{M_2}a_{m,<i-1>}b_{m,k}g_k(i-1-\tau_{m,<i-1>}))\right\|^2$$

$$-3L^2\left(\left\|\sum_{n=i-1-\tau_{max}}^{n=i-1}\frac{1}{K}\bar{\eta}(\sum_{k=1}^{K_1}\sum_{m=1}^{M_1}a_{m,<i-1>}b_{m,k}(g_k(n-1-\tau_{m,<n-1>}) - g(n-1-\tau_{m,<n-1>}))\right.\right.$$

$$\left.+\sum_{k=1}^{K_2}\sum_{m=1}^{M_2}a_{m,<i-1>}b_{m,k}(g_k(n-1-\tau_{m,<n-1>}) - g(n-1-\tau_{m,<n-1>})))\right\|^2 + \left\|\sum_{n=i-1-\tau_{max}}^{n=i-1}\frac{1}{K}\bar{\eta}\sum_{k=1}^{K}\sum_{m=1}^{M}a_{m,<i-1>}b_{m,k}g(n-1-\tau_{m,<n-1>})\right\|^2\right)$$

$$+\frac{3L\bar{\eta}^2}{2K^2}(\sum_{k=1}^{K_1}\sum_{m=1}^{M_1}a_{m,<i-1>}^2 b_{m,k}^2\varsigma^2 + \sum_{k=1}^{K_2}\sum_{m=1}^{M_2}a_{m,<i-1>}^2 b_{m,k}^2\varsigma^2) + \frac{3L\bar{\eta}^2}{2}\left\|\frac{1}{K}\sum_{k=1}^{K}\sum_{m=1}^{M}g(i-1-\tau_{m,<i-1>})\right\|^2$$

$$\leq -\frac{1}{2}\bar{\eta}(\|g(w_{<i-1>})\|^2 + \left\|\frac{1}{K}(\sum_{k=1}^{K_1}\sum_{m=1}^{M_1}a_{m,<i-1>}b_{m,k}g_k(i-1-\tau_{m,<i-1>}) + \sum_{k=1}^{K_2}\sum_{m=1}^{M_2}a_{m,<i-1>}b_{m,k}g_k(i-1-\tau_{m,<i-1>}))\right\|^2$$

$$-3L^2\left(\left\|\sum_{n=i-1-\tau_{max}}^{n=i-1}\frac{1}{K^2}\bar{\eta}^2(\sum_{k=1}^{K_1}\sum_{m=1}^{M_1}a_{m,<n>}^2 b_{m,k}^2(\|g_k(n-1-\tau_{m,<n-1>}) - g(n-1-\tau_{m,<n-1>})\|^2\right.\right.$$

$$\left.+\sum_{k=1}^{K_2}\sum_{m=1}^{M_2}a_{m,<n>}^2 b_{m,k}^2(\|g_k(n-1-\tau_{m,<n-1>}) - g(n-1-\tau_{m,<n-1>})\|^2) + \left\|\sum_{n=i-1-\tau_{max}}^{n=i-1}\frac{1}{K}\bar{\eta}\sum_{k=1}^{K}\sum_{m=1}^{M}a_{m,<i-1>}b_{m,k}g(n-1-\tau_{m,<n-1>})\right\|^2\right)$$

$$+\frac{3L\bar{\eta}^2}{2K^2}(\sum_{k=1}^{K_1}\sum_{m=1}^{M_1}a_{m,<i-1>}^2 b_{m,k}^2\varsigma^2 + \sum_{k=1}^{K_2}\sum_{m=1}^{M_2}a_{m,<i-1>}^2 b_{m,k}^2\varsigma^2) + \frac{3L\bar{\eta}^2}{2}\left\|\frac{1}{K}\sum_{k=1}^{K}\sum_{m=1}^{M}g(i-1-\tau_{m,<i-1>})\right\|^2$$

$$\leq -\frac{1}{2}\bar{\eta}(\|g(w_{<i-1>})\|^2 + \left\|\frac{1}{K}(\sum_{k=1}^{K_1}\sum_{m=1}^{M_1}a_{m,<i-1>}b_{m,k}g_k(i-1-\tau_{m,<i-1>}) + \sum_{k=1}^{K_2}\sum_{m=1}^{M_2}a_{m,<i-1>}b_{m,k}g_k(i-1-\tau_{m,<i-1>}))\right\|^2$$

$$-3L^2\left(\frac{1}{K^2}\bar{\eta}^2\sum_{n=i-1-\tau_{max}}^{n=i-1}(\sum_{k=1}^{K_1}\sum_{m=1}^{M_1}a_{m,<n>}^2b_{m,k}^2\varsigma^2 + \sum_{k=1}^{K_2}\sum_{m=1}^{M_2}a_{m,<n>}^2b_{m,k}^2\varsigma^2) + \left\|\sum_{n=i-1-\tau_{max}}^{n=i-1}\frac{1}{K}\bar{\eta}\sum_{k=1}^{K}\sum_{m=1}^{M}a_{m,<i-1>}b_{m,k}g(n-1-\tau_{m,<n-1>})\right\|^2\right)$$

$$+\frac{3L\bar{\eta}^2}{2K^2}(\sum_{k=1}^{K_1}\sum_{m=1}^{M_1}a_{m,<i-1>}^2b_{m,k}^2\varsigma^2 + \sum_{k=1}^{K_2}\sum_{m=1}^{M_2}a_{m,<i-1>}^2b_{m,k}^2\varsigma^2) + \frac{3L\bar{\eta}^2}{2}\left\|\frac{1}{K}\sum_{k=1}^{K}\sum_{m=1}^{M}g(i-1-\tau_{m,<i-1>})\right\|^2$$

$$\leq -\frac{1}{2}\bar{\eta}(\|g(w_{<i-1>})\|^2 + \left\|\frac{1}{K}(\sum_{k=1}^{K_1}\sum_{m=1}^{M_1}a_{m,<i-1>}b_{m,k}g_k(i-1-\tau_{m,<i-1>}) + \sum_{k=1}^{K_2}\sum_{m=1}^{M_2}a_{m,<i-1>}b_{m,k}g_k(i-1-\tau_{m,<i-1>}))\right\|^2$$

$$-3L^2\left(\frac{1}{K^2}\bar{\eta}^2\sum_{n=i-1-\tau_{max}}^{n=i-1}(\sum_{k=1}^{K_1}\sum_{m=1}^{M_1}a_{m,<n>}^2b_{m,k}^2\varsigma^2 + \sum_{k=1}^{K_2}\sum_{m=1}^{M_2}a_{m,<n>}^2b_{m,k}^2\varsigma^2) + \tau_{max}\bar{\eta}^2\sum_{n=i-1-\tau_{max}}^{n=i-1}\left\|\frac{1}{K}\sum_{k=1}^{K}\sum_{m=1}^{M}a_{m,<i-1>}b_{m,k}g(n-1-\tau_{m,<n-1>})\right\|^2\right)$$

$$+\frac{3L\bar{\eta}^2}{2K^2}(\sum_{k=1}^{K_1}\sum_{m=1}^{M_1}a_{m,<i-1>}^2b_{m,k}^2\varsigma^2 + \sum_{k=1}^{K_2}\sum_{m=1}^{M_2}a_{m,<i-1>}^2b_{m,k}^2\varsigma^2) + \frac{3L\bar{\eta}^2}{2}\left\|\frac{1}{K}\sum_{k=1}^{K}\sum_{m=1}^{M}g(i-1-\tau_{m,<i-1>})\right\|^2$$

$$\leq -\frac{1}{2}\bar{\eta}(\|g(w_{<i-1>})\|^2 - \frac{1}{2}\bar{\eta}\left\|\frac{1}{K}(\sum_{k=1}^{K_1}\sum_{m=1}^{M_1}a_{m,<i-1>}b_{m,k}g_k(i-1-\tau_{m,<i-1>}) + \sum_{k=1}^{K_2}\sum_{m=1}^{M_2}a_{m,<i-1>}b_{m,k}g_k(i-1-\tau_{m,<i-1>}))\right\|^2$$

$$+\frac{3}{2}\frac{L^2\bar{\eta}^3}{K^2}(\sum_{n=i-1-\tau_{max}}^{n=i-1}(\sum_{k=1}^{K_1}\sum_{m=1}^{M_1}a_{m,<n>}^2b_{m,k}^2\varsigma^2 + \sum_{k=1}^{K_2}\sum_{m=1}^{M_2}a_{m,<n>}^2b_{m,k}^2\varsigma^2) + \frac{3}{2}L^2\tau_{max}\bar{\eta}^3\sum_{n=i-1-\tau_{max}}^{n=i-1}\left\|\frac{1}{K}\sum_{k=1}^{K}\sum_{m=1}^{M}a_{m,<i-1>}b_{m,k}g(n-1-\tau_{m,<n-1>})\right\|^2$$

$$+\frac{3L\bar{\eta}^2}{2K^2}(\sum_{k=1}^{K_1}\sum_{m=1}^{M_1}a_{m,<i-1>}^2b_{m,k}^2\varsigma^2 + \sum_{k=1}^{K_2}\sum_{m=1}^{M_2}a_{m,<i-1>}^2b_{m,k}^2\varsigma^2) + \frac{3L\bar{\eta}^2}{2}\left\|\frac{1}{K}\sum_{k=1}^{K}\sum_{m=1}^{M}g(i-1-\tau_{m,<i-1>})\right\|^2$$

$$= -\frac{1}{2}\bar{\eta}(\|g(w_{<i-1>})\|^2 + (\frac{3L\bar{\eta}^2}{2} - \frac{1}{2}\bar{\eta})\left\|\frac{1}{K}\sum_{k=1}^{K}\sum_{m=1}^{M}g(i-1-\tau_{m,<i-1>})\right\|^2$$

$$+\frac{3}{2}L^2\tau_{max}\bar{\eta}^3\sum_{n=i-1-\tau_{max}}^{n=i-1}\left\|\frac{1}{K}\sum_{k=1}^{K}\sum_{m=1}^{M}a_{m,<i-1>}b_{m,k}g(n-1-\tau_{m,<n-1>})\right\|^2$$

$$+(\frac{3L\bar{\eta}^3}{2K^2}(\sum_{k=1}^{K_1}\sum_{m=1}^{M_1}a_{m,<i-1>}^2b_{m,k}^2 + \sum_{k=1}^{K_2}\sum_{m=1}^{M_2}a_{m,<i-1>}^2b_{m,k}^2) + \frac{3L\bar{\eta}^2}{2K^2}(\sum_{k=1}^{K_1}\sum_{m=1}^{M_1}a_{m,<i-1>}^2b_{m,k}^2 + \sum_{k=1}^{K_2}\sum_{m=1}^{M_2}a_{m,<i-1>}^2b_{m,k}^2))\varsigma^2$$

$$\mathrm{E}\left[F(w_{<P+1>})\right] - F(w_{<1>})$$

$$= -\sum_{l=1}^{P}\frac{1}{2}\bar{\eta}(\|g(w_{<l>})\|^2 + \sum_{l=1}^{P}(\frac{3L\bar{\eta}^2}{2} - \frac{1}{2}\bar{\eta})\left\|\frac{1}{K}\sum_{k=1}^{K}\sum_{m=1}^{M}g(i-1-\tau_{m,<i-1>})\right\|^2$$

$$+\sum_{l=1}^{P}\frac{3}{2}L^2\tau_{max}\bar{\eta}^3\sum_{n=i-1-\tau_{max}}^{n=i-1}\left\|\frac{1}{K}\sum_{k=1}^{K}\sum_{m=1}^{M}a_{m,<i-1>}b_{m,k}g(n-1-\tau_{m,<n-1>})\right\|^2$$

$$+\sum_{l=1}^{P}(\frac{3L\bar{\eta}^3}{2K^2}(\sum_{k=1}^{K_1}\sum_{m=1}^{M_1}a_{m,<i-1>}^2b_{m,k}^2 + \sum_{k=1}^{K_2}\sum_{m=1}^{M_2}a_{m,<i-1>}^2b_{m,k}^2) + \frac{3L\bar{\eta}^2}{2K^2}(\sum_{k=1}^{K_1}\sum_{m=1}^{M_1}a_{m,<i-1>}^2b_{m,k}^2 + \sum_{k=1}^{K_2}\sum_{m=1}^{M_2}a_{m,<i-1>}^2b_{m,k}^2))\varsigma^2$$

$$= -\sum_{l=1}^{P}\frac{1}{2}\bar{\eta}(\|g(w_{<l>})\|^2 + (\frac{3}{2}\bar{\eta}^2(L + L^2\tau_{max}\bar{\eta}^2) - \frac{1}{2}\bar{\eta})\sum_{l=1}^{P}\left\|\frac{1}{K}\sum_{k=1}^{K}\sum_{m=1}^{M}a_{m,<t-1>}b_{m,k}g(t-1-\tau_{m,<l>})\right\|^2$$

$$+(\frac{3L\bar{\eta}^3}{2K^2} + \frac{3L\bar{\eta}^2\tau_{max}}{2K^2})\sum_{l=1}^{P}\sum_{k=1}^{K_1}\sum_{m=1}^{M_1}a_{m,<l>}^2b_{m,k}^2 + \sum_{l=1}^{P}\sum_{k=1}^{K_2}\sum_{m=1}^{M_2}a_{m,<l>}^2b_{m,k}^2)\varsigma^2$$

$$\leq -\sum_{l=1}^{P}\frac{1}{2}\bar{\eta}(\|g(w_{<l>})\|^2 + (\frac{3L\bar{\eta}^3}{2K^2} + \frac{3L\bar{\eta}^2\tau_{max}}{2K^2})(\sum_{l=1}^{P}\sum_{k=1}^{K_1}\sum_{m=1}^{M_1}a_{m,<l>}^2b_{m,k}^2 + \sum_{l=1}^{P}\sum_{k=1}^{K_2}\sum_{m=1}^{M_2}a_{m,<l>}^2b_{m,k}^2)\varsigma^2$$

$$\frac{1}{P}\sum_{l=1}^{P}\|g(w_{<l>})\|^2$$

$$\leq \frac{2}{\bar{\eta}P}(F(w_{<1>}) - F(w_*)) + \frac{3L\bar{\eta}}{K}(1+\bar{\eta}\tau_{max})m(C_1+C_2)\varsigma^2$$

Proof for lemma 4.

$$Q_{m,<i>}(t+1)^2 = \left(Q_{m,<i>}(t) + d_{m,<i>}(t) - c_{m,<i>}(t)\right)^2$$
$$\leq Q_{m,<i>}(t)^2 + d_{m,<i>}(t)^2 + c_{m,<i>}(t)^2 + 2Q_{m,<i>}(t)(d_{m,<i>}(t) - c_{m,<i>}(t))$$

$$H_{m,<i>}(t+1)^2 = \max[H_{m,<i>}(t) + y_{m,<i>}(t) - d_{m,<i>}(t), 0]^2$$
$$\leq H_{m,<i>}(t)^2 + y_{m,<i>}(t)^2 + d_{m,<i>}(t)^2 + 2H_{m,<i>}(t)(y_{m,<i>}(t) - d_{m,<i>}(t))$$

$$E_{m,<i>}(t+1)^2 = (E_{m,<i>}(t) - e_{m\_up,<i>}(t) - e_{m\_com,<i>}(t) + e_{m\_store,<i>}(t))^2$$
$$\leq E_{m,<i>}(t)^2 + (e_{m\_up,<i>}(t) + e_{m\_com,<i>}(t))^2 + e_{m\_store,<i>}(t)^2$$
$$+ 2E_{m,<i>}(t)(e_{m\_store,<i>}(t) - (e_{m\_up,<i>}(t) + e_{m\_com,<i>}(t)))$$

$$R_{m,<i>}(t+1)^2 = \max[R_{m,<i>}(t) - f_m(t), 0]^2$$
$$\leq R_{m,<i>}(t)^2 + f_m(t)^2 - 2R_{m,<i>}(t)f_m(t)$$

$$R_{server,<i>}(t+1)^2 = \left(\max[R_{server,<i>}(t) - F(t), 0] + \sum_{m \in M_{non\_stragglers}} c_{m,<i>}(t)\xi_m\right)^2$$
$$\leq R_{server,<i>}(t)^2 + F(t)^2 + \sum_{m \in M_{non\_stragglers}} (c_{m,<i>}(t)\xi_m)^2$$
$$+ 2R_{server,<i>}(t)(\sum_{m \in M_{non\_stragglers}} c_{m,<i>}(t)\xi_m - F(t))$$

$$\Delta_V(t) = \frac{1}{2} \sum_{m \in M_{non\_stragglers}} (d_{m,<i>}(t)^2 + c_{m,<i>}(t)^2 + y_{m,<i>}(t)^2 + d_{m,<i>}(t)^2 + (e_{m\_up,<i>}(t) + e_{m\_com,<i>}(t))^2$$
$$+ e_{m\_store,<i>}(t)^2 + f_m(t)^2) + \frac{1}{2}(F(t)^2 + \sum_{m \in M_{non\_stragglers}} (c_{m,<i>}(t)\xi_m)^2) - V\mathbf{E}\left\{\sum_{m \in M_{non\_stragglers}} \log(1 + y_m(t)|\Theta(t)\right\}$$
$$+ \sum_{m \in M_{non\_stragglers}} Q_{m,<i>}(t)\mathbf{E}\{(d_{m,<i>}(t) - c_{m,<i>}(t)|\Theta(t)\} + \sum_{m \in M_{non\_stragglers}} H_{m,<i>}(t)\mathbf{E}\{(y_{m,<i>}(t) - d_{m,<i>}(t))|\Theta(t)\}$$
$$+ \sum_{m \in M_{non\_stragglers}} E_{m,<i>}(t)\mathbf{E}\{(e_{m\_store,<i>}(t) - (e_{m\_up,<i>}(t) + e_{m\_com,<i>}(t)))|\Theta(t)\}$$
$$- \sum_{m \in M_{non\_stragglers}} R_{m,<i>}(t)\mathbf{E}\{f_m(t)|\Theta(t)\} + R_{server,<i>}(t)\mathbf{E}\left\{(\sum_{m \in M_{non\_stragglers}} c_{m,<i>}(t)\xi_m - F(t))|\Theta(t)\right\}$$

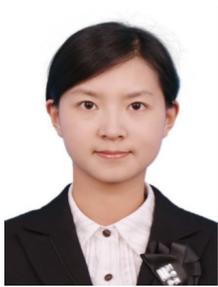 **Tingting Yang** (M'13) received the Ph.D. degrees from Dalian Maritime University, China, in 2010. She is currently a professor in the School of Electrical Engineering and Intelligentization, Dongguan University of Technology, China. Since September 2012, she has been a visiting scholar at the Broadband Communications Research (BBCR) Lab at the Department of Electrical and Computer Engineering, University of Waterloo, Canada. Her research interests are in the areas of maritime wideband communication networks, DTN networks, and green wireless communication. She serves as the associate Editor-in-Chief of the IET Communications, as well as the advisory editor for SpringerPlus. She also serves as a TPC Member for IEEE ICC'14, ICC'15.

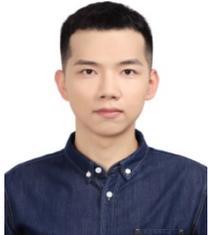 **Xinghan Wang** He is currently pursuing the Ph.D degree with the School of Cyber Science and Engineering, Southeast University, Nanjing, China. His research interest includes network protocol design and analysis, reinforcement learning and edge computing.

He is the elected IEEE Communications Society vice president for Technical & Educational Activities, vice president for Publications, Member-at-Large on the Board of Governors, chair of the Distinguished Lecturer Selection Committee, member of IEEE ComSoc fellow Selection Committee. He was/is the editor-in-chief of the IEEE Internet of Things Journal, IEEE Network, IET Communications, and Peer-to-Peer Networking and Applications